\documentclass[12pt,a4paper]{amsart}

\usepackage{mathtools}
\usepackage{standard}
\usepackage{amsrefs}
\usepackage{enumerate}
\usepackage[usenames,dvipsnames]{xcolor}
\usepackage[colorlinks=true,linkcolor=Red,citecolor=Green]{hyperref}
\usepackage[depth=2]{bookmark}
\usepackage{tikz}

\usepackage[utf8x]{inputenc}
\usepackage[T1]{fontenc}
\usepackage{lmodern}
\usepackage{microtype}
\usepackage[bb=stix]{mathalpha}

\numberwithin{theorem}{section}

\newcommand{\FT}{\mathcal{F}}
\newcommand{\pt}{{}_{\mathrm{pt}}}

\usepackage{listings}

\definecolor{keywordcolor}{rgb}{0.7, 0.1, 0.1}   
\definecolor{commentcolor}{rgb}{0.4, 0.4, 0.4}   
\definecolor{symbolcolor}{rgb}{0.0, 0.1, 0.6}    
\definecolor{sortcolor}{rgb}{0.1, 0.5, 0.1}      
\definecolor{shadecolor}{rgb}{0.98, 0.98, 0.98}  

\usepackage{xspace}
\newcommand{\mathlib}{\texttt{mathlib}\xspace}
\newcommand{\ignore}[2]{#2}

\usepackage{fontawesome}
\newcommand{\link}{\!\ensuremath{{}^\text{\faExternalLink}}\!}
\newcommand{\tlink}[1]{
\href{https://github.com/mcdoll/mathlib4/blob/formal_schwartz_v2/Mathlib/#1}{\link}
}
\newcommand{\prlink}[1]{
\href{https://github.com/leanprover-community/mathlib4/pull/#1}{\link}
}

\title[Formalizing tempered distributions]{
Formalizing Schwartz functions and tempered distributions}

\author[M. Doll]{Moritz Doll}
\address{School of Mathematical and Physical Sciences, Macquarie University \\ NSW 2109 \\ Australia}
\address{School of Mathematics and Statistics, University of Melbourne \\ VIC 3010 \\ Australia}
\email{moritz.doll@unimelb.edu.au}

\begin{document}
\begin{abstract}
Distribution theory is a cornerstone of the theory of partial differential equations.
We report on the progress of formalizing the theory of tempered distributions in the interactive proof
assistant Lean, which is the first formalization in any proof assistant.
We give an overview of the mathematical theory and
highlight key aspects of the formalization that differ from the classical presentation.
As an application, we prove that the Fourier transform extends to a linear isometry on $L^2$ and
we define Sobolev spaces via the Fourier transform on tempered distributions.
\end{abstract}

\maketitle

\section{Introduction}

The theory of distributions plays a major role in the theory of partial differential equations and
more generally in analysis. In this article, we report on the progress of formalizing the
theory of \emph{tempered distributions}, an important subclass of distributions, in the interactive proof assistant Lean~\cite{deMoura2015},
making use of the extensive mathematical library \mathlib\cite{mathlib2020}.

Tempered distributions are typically defined as the topological dual of the space of Schwartz functions $\schwartz(\RR^d)$.
A function $f : \RR^d \to \CC$ is called a \emph{Schwartz function} if it is smooth and for all multi-indices $\alpha, \beta \in \NN_0^d$ there exist $C_{\alpha, \beta} \geq 0$
such that
\begin{align}\label{eq:schwartz}
    \abs{x^\alpha \partial^\beta f(x)} < C_{\alpha,\beta}
\end{align}
for all $x \in \RR^d$.
The Schwartz space is the set of all Schwartz functions and is denoted by $\schwartz(\RR^d)$.
The family of seminorms $p_{\alpha,\beta}(f) \coloneqq \sup_{x \in \RR^d} \abs{x^\alpha \partial^\beta f(x)}$
turns $\schwartz(\RR^d)$ into a locally convex topological vector space.
Hence, it makes sense to consider the topological dual $\schwartz'(\RR^d)$ consisting of all continuous linear functionals $\schwartz(\RR^d) \to \CC$.
The space of tempered distributions contains all $L^p$ functions $\phi$, via the inclusion map $f \mapsto \int_{\RR^d} f(x) \phi(x) \,dx$.
An important example of a tempered distributions that is not a classical function is the Dirac delta $\delta \in \schwartz'(\RR^d)$, defined by $\delta(f) \coloneqq f(0)$.

One of the main reasons to consider the Schwartz space is that the Fourier transform
\begin{align*}
    (\FT u)(\xi) \coloneqq \int_{\RR^d} e^{-2\pi ix \xi} u(x) \, dx
\end{align*}
is an isomorphism on $\schwartz(\RR^d)$.
Using duality, one can extend the definition of the Fourier transform to tempered distributions.
Since the Fourier transform diagonalizes linear constant coefficient differential operators,
one can use the Fourier transform on tempered distributions to solve various differential equations and
prove precise regularity results for the solution.
For a more thorough introduction to distribution theory and its applications in partial differential equations and mathematical physics,
we refer to Hörmander~\cite{Hormander1} and Reed--Simon~\cite{ReedSimon1}.
An example of the use of Schwartz functions outside the field of partial differential equations is the recent resolution of the 8-dimensional sphere packing problem by Viazovska~\cite{Viazovska2017}.
The on-going formalization of Viazovska's result (see \href{https://github.com/thefundamentaltheor3m/Sphere-Packing-Lean}{Sphere-Packing-Lean}) uses the formalization of Schwartz function presented here.

\subsection*{Previous work}
This project heavily relies on the existing formalization of mathematics in \mathlib, in particular
the topology library (originally introduced by Hölzl building on a previous Isabelle/HOL formalization~(cf.~\cite{Holzl2013}) and expanded on by Buzzard--Commelin--Massot~\cite{Buzzard2020}) ,
the calculus library~\cite{Gouezel2025}, and Fourier transform of integrable functions~\cite{Loeffler2025}.

While there has been formalization of solutions to specific partial differential equations using elementary
methods~(cf.~\cite{Deniz2022} in HOL), there has been no formalization of Sobolev spaces in any interactive theorem prover.
We note however that the Gagliardo--Nirenberg--Sobolev inequality has been recently formalized
by van Doorn--Macbeth~\cite{vanDoorn2024} in Lean.

\subsection*{Structure of the article}
The article is structured as follows:
we give a broad overview of the different design choices when formalizing
function spaces, and the key differences between the informal mathematical definitions
and the formalized version.
Then we expand on the theory of locally convex spaces, the Schwartz space, and tempered distribution
indicating the key differences between the usual definitions and the formalization as well as challenges in the formalization.
Finally, we provide two examples that show that the formalization of tempered distributions is practical in sense that it provides a good basis
for formalizing partial differential equations.

\subsection*{Code examples}
The code associated to the formalization can be found at \url{https://github.com/mcdoll/mathlib4/tree/formal_schwartz_v2}, which is a fixed version of \mathlib.
We use links \tlink{Analysis/Fourier/LpSpace.lean\#L49} to refer to specific sections of that repository.
The code snippets in this article might have been slightly modified for improved readability and we refer to the repository
for the formally correct results.

\subsection*{Acknowledgments}
The author is grateful for various conversations with members of the \mathlib community,
in particular, Anatole Dedecker, Heather Macbeth, and Jireh Loreaux.
We would also like to thank S\'ebastien Gou\"ezel for reviewing many pull requests into \mathlib
and we also wish to thank an anonymous referee for helpful suggestions, improving the overall quality
of the paper.

\section{Function spaces in \texorpdfstring{\mathlib}{mathlib}}\label{spaces_in_mathlib}
While a full description of the type theory of Lean is beyond the scope of this article,
we will indicate how formalized mathematics in Lean differs from the more familiar informal writings
and which considerations guide the way how certain mathematical objects are formalized.

Every object in Lean has a type and an object \lstinline{x} being of type \lstinline{X} is denoted by \lstinline{x : X}. The most
important types are the type of all propositions, \lstinline{Prop}, and the type of functions from \lstinline{X} to \lstinline{Y}, \lstinline{X → Y}.

If one wants to define a function with additional properties, then there are broadly speaking two different ways to formalize this, unbundled and bundled types.
Given a property of functions, \lstinline{p : (E → F) → Prop}, an \emph{unbundled function} would be described as
\begin{lstlisting}
variable (f : E → F) (hf : p f) 
\end{lstlisting}
whereas to define the corresponding \emph{bundled function} one defines a structure
\begin{lstlisting}
structure PFunction (E F) where
  toFun : E → F
  property : p toFun
\end{lstlisting}
and the elements \lstinline{f' : PFunction E F} of this new type are exactly functions together with the property \lstinline{p}.
We note that while for \lstinline{x : E} the application \lstinline{f' x} is a-priori not defined, Lean provides
a mechanism so that \lstinline{f' x} is defined as \lstinline{f'.toFun x} (see also \cite{Baanen2022}*{Section~6.3}).

Unbundled functions have the advantage that we do not need to perform any type conversions.
As an example consider the fact that every differentiable function is continuous: if \mathlib used the bundled approach, then
there would be a map from the type of differentiable functions to the type of continuous functions. In this case it is easier
to use unbundled functions, then the same statement is just
\begin{lstlisting}
theorem Differentiable.continuous (f : E → F) (h : Differentiable 𝕜 f) :
  Continuous f
\end{lstlisting}

Since we want to speak of algebraic properties, topologies, etc on the \emph{space of Schwartz functions}, it makes sense to
use the bundled approach for the Schwartz space. Other functions we might consider such as smooth functions with temperate growth
are kept unbundled.
While algebraic properties are possible to express in both the unbundled and bundled approach,
topologies in Lean are defined on a type, so one either has to use a subtype of functions or define a structure as above, where
the later is preferred throughout \mathlib.

One of the features of type theory is that every object has exactly one type and therefore there is no concept of inclusion of types.
Hence it is not possible to state the inclusion of $\schwartz$ in $L^p$ as an inclusion of sets, but we have to define
the map from the type of Schwartz functions to the type of $L^p$ functions and prove the corresponding properties of that map.
\begin{remark}
    We note that strictly speaking $\schwartz(\RR^d) \subset L^p(\RR^d)$ is not entirely correct in set theory either, since $L^p$ functions are defined as
    equivalence classes of measurable functions. This abuse of notation becomes even more obvious for the ``inclusion'' of Schwartz functions
    in the space of tempered distributions, $\schwartz \subset \schwartz'$.
\end{remark}

Moreover, it is convenient to introduce type classes that purely introduce notation, in order to avoid having to explicitly differentiate between say the Fourier transform for
functions, Schwartz functions, and tempered distributions.

\begin{lstlisting}[mathescape]
class FourierTransform (E) (outParam F) where
  fourierTransform : E → F

notation "$\FT$" => FourierTransform.fourierTransform
\end{lstlisting}
So for example if \lstinline{u} is a tempered distribution, then \lstinline{$\FT$ u} denotes the distributional Fourier transform and if \lstinline{u} is a Schwartz function then
\lstinline{$\FT$ u} is the Fourier transform as a Schwartz function.
When using this notation in practice, the type class inference system only knows the type \lstinline{E}, the domain of the Fourier transform, but not the codomain.
Lean provides the \lstinline{outParam} command, that indicates that the parameter \lstinline{F} can be inferred by the type system, so that we do not have to specify
that \lstinline{$\FT$ u} is a tempered distribution. This comes at the minor cost that we can only define one Fourier transform for each domain type.

\section{Locally convex spaces}
There are two very different definitions of locally convex topological vector spaces (LCTVS). Firstly,
a topological vector space $X$ is said to be locally convex if $0$ admits a basis of neighborhoods made of convex sets~\cite{Bourbaki1987}*{p.~II.23}.

On the other hand Reed--Simon~\cite{ReedSimon1}*{Section V.1} define a LCTVS as vector space together with a family of seminorms.
These seminorms induce a topology, which can be shown~\cite{Bourbaki1987}*{p.~II.24} to be locally convex in the topological sense.
For the concrete task of constructing the Schwartz space, it is obvious that the later definition is more useful.
Note that both definitions and their equivalence have been formalized in \mathlib%
(see \tlink{Analysis/LocallyConvex/WithSeminorms.lean\#L918} and
\tlink{Mathlib.Analysis.LocallyConvex.AbsConvexOpen.lean\#L109}.

Recall that a seminorm on a vector space $X$ over a normed field $\mathbb{k}$ is a function $p : X \to \RR$, such that
$p(x + y) \leq p(x) + p(y)$ for all $x, y \in X$ (subadditivity) and $p(r x) = \abs{r} \cdot p(x)$ for all $r \in \mathbb{k}$ and $x \in X$ (absolute homogeneity).

In \mathlib this definition reads as\tlink{Analysis/Seminorm.lean\#L51}
\begin{lstlisting}
structure Seminorm (𝕜 E) [SeminormedRing 𝕜] [AddGroup E] [SMul 𝕜 E] where
  toFun : E → ℝ
  map_zero' : toFun 0 = 0
  add_le' (r s : E) : toFun (r + s) ≤ toFun r + toFun s
  neg' (r : E) : toFun (-r) = toFun r
  smul' (a : 𝕜) (x : E) : toFun (a • x) = ‖a‖ * toFun x
\end{lstlisting}
We note that the assumptions on $X$ and $\mathbb{k}$ are significantly relaxed.
The slightly redundant looking properties
are due to the fact that this definition is a specialization of seminorms on groups. In particular, \mathlib provides
a definition to construct a seminorm by proving only subadditivity and absolute homogeneity.\tlink{Analysis/Seminorm.lean\#L74}

A family of seminorms is now simply defined as a map from any type into \lstinline{Seminorm 𝕜 E}:
\begin{lstlisting}
abbrev SeminormFamily 𝕜 E ι := (ι → Seminorm 𝕜 E)
\end{lstlisting}

The definition of a locally convex topological vector space now proceeds in two steps: first
we define the topology induced by the family of seminorms by describing a neighborhood base at $0 \in E$.
Secondly, we show that if $E$ is a vector space with topology $t$
and $t$ coincides with the induced topology of a family of seminorms, then the topology is
compatible with the linear structure and locally convex.

For a family of seminorms indexed by $A$, the neighborhood base at $0$ is given by
the sets $\set{ N_{\alpha_1,\dotsc,\alpha_n; r} \;|\;\, \alpha_1,\dotsc\alpha_n \in A,\, r > 0}$
where
\begin{align}\label{eq:sup_ball}
    N_{\alpha_1,\dotsc,\alpha_n;r} \coloneqq \set{x \in E \;|\;\, p_{\alpha_i}(x) < r \text{ for all } i \in 1,\dotsc,n}
\end{align}
(see Reed--Simon~\cite{ReedSimon1}*{p.~125}\footnote{Note that Reed--Simon~\cite{ReedSimon1} uses nets, whereas \mathlib's topology library uses filters~(see \cite{Buzzard2020}),
but this is irrelevant here.}).

Noting that the right hand side of \eqref{eq:sup_ball} is just the ball of radius $r$ of the supremum of the seminorms,
we can easily translate this definition as follows:%
\tlink{Analysis/LocallyConvex/WithSeminorms.lean\#L83}
\begin{lstlisting}
def basisSets (p : SeminormFamily 𝕜 E ι) : Set (Set E) :=
  $\bigcup$ (s : Finset ι) (r) (_ : 0 < r), singleton (ball (s.sup p) (0 : E) r)
\end{lstlisting}

We now use general construction, \lstinline{ModuleFilterBasis}, to obtain the topology and derive that
the topology is compatible with the linear structure.
\lstinline{ModuleFilterBasis} was first introduced by Patrick Massot, see~\tlink{Topology/Algebra/FilterBasis.lean}.

In order to prove that $E$ is locally convex, we introduce the predicate \lstinline{WithSeminorms},
which states that the topology on $E$ is induced by a family of seminorms.%
\tlink{Analysis/LocallyConvex/WithSeminorms.lean\#L284}
\begin{lstlisting}
structure WithSeminorms (p : SeminormFamily 𝕜 E ι) [t : TopologicalSpace E] : Prop where
  topology_eq_withSeminorms : t = p.moduleFilterBasis.topology
\end{lstlisting}

\begin{remark}
    This structure has the advantage that the family of seminorms is not part of the definition of the topological space.
    Otherwise Lean would consider the same space with different families of seminorms as
    being different, which is highly undesirable since one might want to work with multiple families of seminorms defining the same topology.
    On the other hand, the predicate \lstinline{WithSeminorms} makes it easy to prove continuity and convergence results
    by proving bounds of the seminorms.

    In the case of the Schwartz space, the equality in \lstinline{WithSeminorms} is satisfied by definition, but this does not have to be the case.
    For example, every normed space satisfies \lstinline{WithSeminorms} with its norm (considered as a seminorm), but topology is not defined
    through \lstinline{ModuleFilterBasis}.
\end{remark}

\begin{remark}
    Another possibility is to define the topology via the distance functions
    associated to each seminorm: for each seminorm $p$, we set $d_p : E \times E \to \RR$ with $d_p(x,y) = p(x - y)$. Then we pick the coarsest topology that is finer than the topology induced by $d_p$ for each seminorm $p$.
    However, we chose the approach of filter bases since it automatically implies that $E$ is a topological vector space,
    i.e., the topology is compatible with the linear structure.
\end{remark}

A linear map $f : E \to F$ between locally convex spaces $E$ and $F$ with families of seminorms $\{p_j\}_{j \in J}$ and $\{q_j\}_{j \in J'}$ is bounded if for every seminorm
$q_j$ there exist a finite set of seminorms $\{p_{j_k}\}_{k=1}^n$ and $C > 0$ such that $p_j(x) \leq C \sup_{k} p_{j_k}(f(x))$.
This is formalized as:
\begin{lstlisting}
def IsBounded (p : ι → Seminorm 𝕜 E) (q : ι' → Seminorm 𝕜₂ F) (f : E →ₛₗ[σ₁₂] F) :=
  ∀ i, ∃ s : Finset ι, ∃ C : ℝ≥0, (q i).comp f ≤ C • s.sup p
\end{lstlisting}
Note that the composition of a linear map and a seminorm is again a seminorm, and therefore we can express the inequality on the level of seminorms.

Then we have the following criterion for continuity:
\tlink{Analysis/LocallyConvex/WithSeminorms.lean\#L671}
\begin{lstlisting}
theorem continuous_of_isBounded (hp : WithSeminorms p) (hq : WithSeminorms q)
    (f : E →ₛₗ[τ₁₂] F) (hf : Seminorm.IsBounded p q f) : Continuous f
\end{lstlisting}

In order to unify the continuity statements to also contain the case of normed spaces or topological vector spaces that
are not necessarily locally convex, we introduce the notion of \emph{von Neumann bounded sets}~\tlink{Analysis/LocallyConvex/Bounded.lean}
(see Reed--Simon~\cite{ReedSimon1}*{p.~165} for a definition) and prove that for spaces where the topology is induced by a family of seminorms,
boundedness can be expressed in terms of the seminorms.%
\tlink{Analysis/LocallyConvex/WithSeminorms.lean\#L461}
\begin{lstlisting}
theorem WithSeminorms.isVonNBounded_iff_finset_seminorm_bounded {s : Set E}
    (hp : WithSeminorms p) :
  IsVonNBounded 𝕜 s ↔ ∀ I : Finset ι, ∃ r > 0, ∀ x ∈ s, I.sup p x < r
\end{lstlisting}

If the family of seminorms $\{p_j\}$ is countable, then $E$ is first countable%
\tlink{Analysis/LocallyConvex/WithSeminorms.lean\#L1070}
and the continuity follows from the abstract result:%
\tlink{Analysis/LocallyConvex/ContinuousOfBounded.lean\#L123}
\begin{lstlisting}
theorem LinearMap.continuous_of_locally_bounded [FirstCountableTopology E]
    (f : E →ₗ[𝕜] F) (hf : ∀ s, IsVonNBounded 𝕜 s → IsVonNBounded 𝕜 (f '' s)) :
  Continuous f
\end{lstlisting}

\section{Schwartz space}

Throughout this section, we will assume that $E, F$ are normed spaces over the reals, which in \mathlib is phrased as
\begin{lstlisting}
variable (E) [NormedAddCommGroup E] [NormedSpace ℝ E]
\end{lstlisting}

The formal definition of the Schwartz space is the following:%
\tlink{Analysis/Distribution/SchwartzSpace/Basic.lean\#L78}
\begin{lstlisting}[mathescape]
structure SchwartzMap (E F) where
  toFun : E → F
  smooth' : ContDiff ℝ ∞ toFun
  decay' (k n : ℕ) : ∃ (C : ℝ), ∀ (x : E), ‖x‖ ^ k * ‖iteratedFDeriv ℝ n toFun x‖ ≤ C
\end{lstlisting}
We also introduce the notation
\begin{lstlisting}[mathescape]
    $\schwartz$(E, F) := SchwartzMap E F
\end{lstlisting}
\begin{remark}
    We note a few differences to the usual definition~\eqref{eq:schwartz}:
    \begin{enumerate}
        \item We allow for normed vector space $E$ and $F$ instead of the domain being $\RR^d$ and the codomain being the complex numbers $\CC$.
        \item We avoid using a basis in $E$ to express both the iterated derivatives as well as the polynomial bound.
    \end{enumerate}
    Both generalizations are necessary to obtain a useful formalization.
    In particular, note that $\RR^d$ for $d = 1$ is isomorphic, but not equal to $\RR$, so using $\RR^d$ in the definition of $\schwartz$
    would cause a lot of friction already for the case $d = 1$. Similarly, the cartesian product of $\RR^d$ and $\RR^{d'}$ is not definitionally equal to $\RR^{d+d'}$.
    Moreover, \mathlib uses type synonyms to describe spaces that are isomorphic (but not canonically) to $\RR^n$.

    Note that here we have not assume that $E$ is finite dimensional or an inner product space. We will make these assumptions whenever they are necessary.

    On the other hand having vector-valued Schwartz functions is important. For example
    the Fr\'echet derivative in \mathlib is defined as a map \lstinline{E → (E →L[𝕜] F)}, meaning a map into the space of continuous linear maps from \lstinline{E} to \lstinline{F},
    and therefore to state that the derivative of a Schwartz function defines a Schwartz function, the target $F$ has to be sufficiently general.
\end{remark}

We also define seminorms \lstinline{SchwartzMap.seminorm (k n : ℕ)},%
\tlink{Analysis/Distribution/SchwartzSpace/Basic.lean\#L417} which are given by
the infimum of all positive \lstinline{c : ℝ} such that 
\lstinline{‖x‖ ^ k * ‖iteratedFDeriv ℝ n f x‖ ≤ c}. The two important consequences are the following two
propositions:
\begin{lstlisting}[mathescape]
theorem seminorm_le_bound (k n : ℕ) (f : $\schwartz$(E, F)) {M : ℝ} (hMp : 0 ≤ M)
    (hM : ∀ x, ‖x‖ ^ k * ‖iteratedFDeriv ℝ n f x‖ ≤ M) :
  SchwartzMap.seminorm 𝕜 k n f ≤ M

theorem le_seminorm (k n : ℕ) (f : $\schwartz$(E, F)) (x : E) :
  ‖x‖ ^ k * ‖iteratedFDeriv ℝ n f x‖ ≤ SchwartzMap.seminorm 𝕜 k n f
\end{lstlisting}
This means that the seminorms \lstinline{SchwartzMap.seminorm 𝕜 k n f} is the supremum of \lstinline{‖x‖ ^ k * ‖iteratedFDeriv ℝ n f x‖}.

Using the \lstinline{WithSeminorms} construction described in the previous section, we can equip with the structure of a
locally convex topological vector space.%
\tlink{Analysis/Distribution/SchwartzSpace/Basic.lean\#L523}

In order to prove that Schwartz functions are dense in $L^p$, we need to show that
every smooth compactly supported function is a Schwartz function.%
\tlink{Analysis/Distribution/SchwartzSpace/Basic.lean\#L562}
\begin{lstlisting}[mathescape]
def HasCompactSupport.toSchwartzMap {f : E → F} (h₁ : HasCompactSupport f)
  (h₂ : ContDiff ℝ ∞ f) : $\schwartz$(E, F)

example {f : E → F} (h₁ : HasCompactSupport f) (h₂ : ContDiff ℝ ∞ f) (x : E) :
  h₁.toSchwartzMap h₂ x = f x
\end{lstlisting}

\subsection{Construction of continuous linear maps on \texorpdfstring{$\schwartz$}{S}}
In order to define operators on $\schwartz$, we introduce an auxiliary construction.%
\tlink{Analysis/Distribution/SchwartzSpace/Basic.lean\#L610}

\begin{lstlisting}[mathescape]
def mkCLM (A : $\schwartz$(D, E) → F → G)
    (hadd : ∀ (f g : $\schwartz$(D, E)) (x), A (f + g) x = A f x + A g x)
    (hsmul : ∀ (a : 𝕜) (f : $\schwartz$(D, E)) (x), A (a • f) x = a • A f x)
    (hsmooth : ∀ f : $\schwartz$(D, E), ContDiff ℝ ∞ (A f))
    (hbound : ∀ n : ℕ × ℕ, ∃ (s : Finset (ℕ × ℕ)) (C : ℝ), 0 ≤ C ∧ ∀ (f : $\schwartz$(D, E)) (x : F), ‖x‖ ^ n.fst * ‖iteratedFDeriv ℝ n.snd (A f) x‖ ≤ C * s.sup
      (schwartzSeminormFamily 𝕜 D E) f) :
  $\schwartz$(D, E) →L[𝕜] $\schwartz$(F, G)
\end{lstlisting}
In short to obtain a continuous linear map $A : \schwartz \to \schwartz$ one has to check that the map is linear, for every $f \in \schwartz$, $A f$ is smooth,
and for every $f \in \schwartz$, we can bound every seminorm by a finite number of seminorms.
In particular, the same estimate of seminorms implies that \lstinline{A f} is a Schwartz function and that the map \lstinline{A} is continuous.

We note that this function is not total, but since we only care about \lstinline{mkCLM} for concrete constructions, this causes no problems
(cf.~\cite{DeLo2025} for a discussion on why non-total functions are usually problematic in dependent type theory).
\begin{remark}
    In fact the actual definition in \mathlib is slightly more general, in that it makes it possible to construct semilinear continuous maps using
    the work of Dupuis--Lewis--Macbeth~\cite{Dupuis2024}.
\end{remark}

This construction made it rather straightforward to define various
operations on $\schwartz(\RR^d)$ and maps from $\schwartz(\RR^d)$ to arbitrary normed spaces (using a similar construction, see%
\tlink{Analysis/Distribution/SchwartzSpace.Basic.lean\#L628}
).
These include differentiation, integration, multiplication by smooth functions of temperate growth, composition with polynomially growing function,
inclusion into bounded continuous functions (contributed by the author), 
inclusion into $L^p$ spaces (contributed by Jack Valmadre~\prlink{20423}), and the Fourier transform (contributed by S\'ebastien Gou\"ezel~\prlink{12144}).

We note that especially for multiplication, composition, and the Fourier transform these constructions provided a test for the robustness of the calculus library of \mathlib \cite{Gouezel2025}.

\section{Tempered distributions}

Now we turn to the definition of tempered distributions $\schwartz'(\RR^d)$.
The first idea to formalize tempered distributions would be as continuous linear maps from $\schwartz(\RR^d)$ to $\CC$:
\begin{lstlisting}[mathescape]
abbrev TemperedDistribution E := $\schwartz$(E, $\CC$) →L[$\CC$] $\CC$
\end{lstlisting}
This definitions has to issues: the we want to include vector-valued distributions and the topology on continuous linear maps is too strong for most applications.

There are two possibilities in the case of vector-valued distributions.
The first option is to declare that tempered distributions should be a dual space, hence a possible candidate for $\schwartz'(\RR^d, F)$ is $\schwartz(\RR^d, F')'$, where
$F'$ denotes the topological dual. Then
the natural inclusion of Schwartz functions into tempered distributions is defined as
\begin{align*}
    \iota : \schwartz(\RR^d, F) &\to \schwartz(\RR^d, F')'\,,\\
    u &\mapsto \left(\phi \mapsto \int \phi(x)(u(x))\, dx \right)\,.
\end{align*}
The second choice, following Schwartz~\cite{Schwartz1957} and Treves~\cite{Treves1967}*{p.~534}, is to define the space of tempered distributions as
the set of continuous linear maps from $\schwartz(\RR^d, \CC)$ to $F$, then the natural inclusion takes the form
\begin{align*}
    \iota : \schwartz(\RR^d, F) &\to \schwartz'(\RR^d, F)\,,\\
    u &\mapsto \left(\phi \mapsto \int \phi(x) \cdot u(x)\, dx \right)\,.
\end{align*}
We have (cf.~\cite{Treves1967}*{p.~534}) that $\schwartz'(\RR^d, F') \cong \schwartz(\RR^d, F)'$, hence if $F$ is not reflexive, the two definitions are not equivalent.
The second definition also has a characterization in terms of the completed tensor product, $\schwartz'(\RR^d, F) \cong \schwartz'(\RR^d) \hat\otimes F$ and the Schwartz kernel
theorem generalizes to this definition of tempered distributions.

Now we turn to the question of the topology.
There are two choices that are almost equivalent: Schwartz~\cite{Schwartz1966} and Treves~\cite{Treves1967} use the topology of bounded convergence (also known as the strong topology), whereas 
Hörmander~\cite{Hormander1}*{p.~164} and Reed--Simon~\cite{ReedSimon1}*{p.~134} use the topology of pointwise convergence,
which is also referred to as the topology of simple convergence~\cite{Bourbaki1987}*{p.~III.14}, the weak topology, or the strong operator topology.

The strong dual of a Montel space is again a Montel space and hence a barrelled space~\cite{Bourbaki1987}*{p.~IV.19}.
Using that $\schwartz$ is a Montel space, we obtain that if $\schwartz'$ is equipped with the topology of \emph{bounded} convergence,
we may apply the Banach--Steinhaus theorem to maps from $\schwartz'$ into any locally convex topological vector space~\cite{Bourbaki1987}*{p.~III.25}.
Moreover, the fact that $\schwartz'(\RR^d)$ is a Montel space, implies that the topology of bounded convergence and the topology of pointwise convergence have the same convergent sequences on $\schwartz'$
even though they are not equal~\cite{Treves1967}*{Section~34.4}.
Since for applications in partial differential equations, it is crucial to check convergence of sequences pointwise, we choose to deviate from Schwartz
and define the space of tempered distributions equipped with the pointwise convergence topology.

Both the bounded convergence topology and the pointwise convergence arise as special cases of the $\mathfrak{S}$-topology,
see Bourbaki~\cite{Bourbaki1987}*{p.~III.13}, and the general construction was contributed to \mathlib by Anatole Dedecker~\prlink{tba}.

The space of bounded linear maps equipped with the pointwise convergence topology is an abbreviation for the $\mathfrak{S}$-topology with $\mathfrak{S} = \{S | S \text{ is a finite set}\}$.%
\tlink{Topology/Algebra/Module/Spaces/PointwiseConvergenceCLM.lean\#L60}
\begin{lstlisting}[mathescape]
abbrev PointwiseConvergenceCLM := UniformConvergenceCLM σ F {s | Finite s}
\end{lstlisting}
We use the notation \lstinline{E →SL$\pt$[σ] F} for \lstinline{PointwiseConvergenceCLM σ E F} and \lstinline{E →L$\pt$[𝕜] F} in the case
of a bounded \lstinline{𝕜}-linear map.

The defining feature of the pointwise convergence topology is that a map is continuous if it is pointwise continuous:%
\tlink{Topology/Algebra/Module/Spaces/PointwiseConvergenceCLM.lean\#L127}
\begin{lstlisting}[mathescape]
theorem continuous_of_continuous_eval {g : α → E →SL$\pt$[σ] F}
    (h : ∀ x, Continuous (g · x)) : Continuous g
\end{lstlisting}

Moreover, using the fact that the family of seminorms $p_u : A \mapsto \norm{A u}$ for $u \in E$ induces the topology,
we obtain the typical characterization of convergence of sequences (phrased here in terms of filters):%
\tlink{Analysis/LocallyConvex/PointwiseConvergenceCLM.lean\#L75}
\begin{lstlisting}[mathescape]
theorem tendsto_nhds {f : Filter α} (u : α → E →SL$\pt$[σ] F) (y₀ : E →SL$\pt$[σ] F) :
    Tendsto u f (𝓝 y₀) ↔ ∀ (x : E) (ε : ℝ), 0 < ε → ∀ᶠ (k : α) in f, ‖u k x - y₀ x‖ < ε

theorem tendsto_nhds_atTop [SemilatticeSup α] [Nonempty α] (u : α → E →SL$\pt$[σ] F)
    (y₀ : E →SL$\pt$[σ] F) :
    Tendsto u atTop (𝓝 y₀) ↔
      ∀ (x : E) (ε : ℝ), 0 < ε → ∃ (k₀ : α), ∀ (k : α), k₀ ≤ k → ‖u k x - y₀ x‖ < ε
\end{lstlisting}

The fact that the pointwise convergence topology is weaker than bounded convergence topology is stated as the fact
that the identity considered as a map from \lstinline{E →SL[σ] F} to \lstinline{E →SL$\pt$[σ] F} is continuous.
Moreover, we provide a continuous linear equivalence between \lstinline{E →L$\pt$[𝕜] 𝕜} and \lstinline{WeakDual 𝕜 E},
which shows that these two topologies coincide.
\tlink{Topology/Algebra/Module/Spaces/PointwiseConvergenceCLM.lean\#L161}

As in the case of continuous linear maps between Schwartz spaces, we provide a function \lstinline{PointwiseConvergenceCLM.mkCLM}
that constructs a linear operator $A$ that is continuous with respect to the pointwise convergence topology:%
\tlink{Analysis/LocallyConvex/PointwiseConvergenceCLM.lean\#L98}
\begin{lstlisting}[mathescape]
def mkCLM (A : (E →SL[σ] F) →ₗ[𝕜₂] D →SL[τ] G)
  (hbound : ∀ (f : D), ∃ (s : Finset E) (C : ℝ≥0),
  ∀ (B : E →SL[σ] F), ∃ (g : E) (_hb : g ∈ s), ‖(A B) f‖ ≤ C • ‖B g‖) :
  (E →SL$\pt$[σ] F) →L[𝕜₂] D →SL$\pt$[τ] G
\end{lstlisting}
Given a continuous linear map $A$ from $E$ to $F$, the adjoint $A^T$ is defined via
\begin{align*}
    A^T(u)(\phi) = u(A \phi)\,,
\end{align*}
where $u$ is a continuous linear map from $F$ to $G$ and $\phi \in E$.
Using the more suggestive notation for the duality pairing $\ang{u, \phi} = u(\phi)$, we have $\ang{A^T u, \phi} = \ang{u, A \phi}$.
This construction works for any $\mathfrak{S}$-topology on continuous linear maps, and in the case of the pointwise convergence topology
takes the form%
\tlink{Topology/Algebra/Module/Spaces/PointwiseConvergenceCLM.lean\#L135}
\begin{lstlisting}[mathescape]
def precomp (L : E →SL[σ] F) : (F →SL$\pt$[τ] G) →L[𝕜₃] E →SL$\pt$[ρ] G
\end{lstlisting}

Now, the definition of tempered distributions is simply an abbreviation for the continuous linear maps into a normed space $V$:%
\tlink{Analysis/Distribution/TemperedDistribution.lean\#L53}
\begin{lstlisting}[mathescape]
abbrev TemperedDistribution E F := $\schwartz$(E, $\CC$) →L$\pt$[$\CC$] V
\end{lstlisting}
with the notation
\begin{lstlisting}[mathescape]
$\schwartz'$(E, F) := TemperedDistribution E F
\end{lstlisting}
We show that the space of Schwartz functions and the space of $L^p$ function continuously embeds into $\schwartz'$:%
\tlink{Analysis/Distribution/TemperedDistribution.lean\#L194}
\begin{lstlisting}[mathescape]
def toTemperedDistributionCLM {p : ℝ≥0∞} (hp : 1 ≤ p) :
    Lp F p μ →L[𝕜] $\schwartz$'(E, F)
\end{lstlisting}
For the embedding of $\schwartz$ into $\schwartz'$, we can use coercions to declutter the notation:%
\tlink{Analysis/Distribution/TemperedDistribution.lean\#L139}
\begin{lstlisting}[mathescape]
variable (f : $\schwartz$(E, F))
#check (f : $\schwartz$'(E, F))
\end{lstlisting}
Naturally, Lean chains coercions together, meaning if we have a coercion $\iota_A : A \to B$ and a coercion $\iota_B : B \to C$, then for \lstinline{a : A}, the
expression \lstinline{a : C} evaluates automatically to $\iota_B (\iota_A a)$. This chaining is performed from the right, in the sense that first $\iota_B$ is found and then
$\iota_A$. Hence, for a coercion to be possible, type $C$ must determine type $B$ uniquely. This is for instance not the case for the coercion \lstinline{Lp F p μ} to \lstinline{$\schwartz$'(E, F)}.
Here, Lean cannot infer the number $p$ from \lstinline{$\schwartz$'(E, F)} (every $1 \leq p < \infty$ is a possible choice). 
Lean does provides the class \lstinline{CoeHead} that can only be the beginning of a coercion chain and this class makes it possible to write the coercion from $L^p$-functions to tempered distributions.%
\tlink{Analysis/Distribution/TemperedDistribution.lean\#L178}
\begin{lstlisting}[mathescape]
variable (f : Lp F p μ)
#check (f : $\schwartz$'(E, F))
\end{lstlisting}

Now we describe how to extend the Fourier transform from Schwartz functions to tempered distributions.

The Fourier transform is symmetric, meaning it satisfies
\begin{align*}
    \ang{\FT f, g} = \ang{f, \FT g}
\end{align*}
for all $f, g \in \schwartz(\RR)$,
where $\ang{f, g} \coloneqq \int f(x) g(x) \, dx$.
This is a direct consequence of Fubini's theorem, since both sides are equal to
\begin{align*}
    \int \int e^{-2\pi i x \xi} f(x) g(\xi) \, dx \, d\xi\,.
\end{align*}

The embedding $\schwartz \to \schwartz'$ is given by $f \mapsto M_f$ with $M_f(g) = \ang{f, g}$.
Hence, we are led to define $\FT : \schwartz' \to \schwartz'$ via $(\FT u)(f) = u(\FT f)$.%
\tlink{Analysis/Distribution/TemperedDistribution.lean\#L469}
Then, it is easy to check that $\FT M_f = M_{\FT f}$.%
\tlink{Analysis/Distribution/TemperedDistribution.lean\#L531}
\begin{lstlisting}[mathescape]
theorem fourier_toTemperedDistributionCLM_eq (f : $\schwartz$(E, F)) :
    $\FT$ (f : $\schwartz$'(E, F)) = $\FT$ f
\end{lstlisting}
Here Lean used automatic coercions on the right hand side \lstinline{$\FT$ f : $\schwartz$'(E, F)}.

Similar arguments can be used to extend differentiation\tlink{Analysis/Distribution/TemperedDistribution.lean\#L362} and multiplication with a function of temperate growth%
\tlink{Analysis/Distribution/TemperedDistribution.lean\#L247} to tempered distributions.

\section{Applications}

\subsection{The Fourier transform on \texorpdfstring{$L^2$}{L2}}
While we defined the Fourier transform for an arbitrary tempered distribution, it is often necessary
to know mapping properties on smaller spaces, such as $L^p$ spaces. One of the most important
results is that the Fourier transform is an isometry on $L^2$.
Note that while this results holds for Abelian locally compact groups, where the Fourier transform is a map from the group to its Pontryagin dual.
The formalization presented here does not generalize to this setting.

To prove this result, we prove first \emph{Plancherel's theorem} stating that  Fourier transform of $f \in \schwartz$ preserves the $L^2$-norm,
$\norm{\FT f}_{L^2}^2 = \norm{f}_{L^2}^2$,
which follows from
\begin{align*}
    \int (\FT f)(x) g(x) \, dx = \int f(x) \overline{(\FT g)(x)} \,dx\,.
\end{align*}
This equality follows from almost the same arguments as the symmetry of the Fourier transform.
\tlink{Analysis/Distribution/SchwartzSpace/Fourier.lean\#L333}
\begin{lstlisting}[mathescape]
theorem norm_fourier_toL2_eq (f : $\schwartz$(E, F)) : ‖($\FT$ f).toLp 2‖ = ‖f.toLp 2‖
\end{lstlisting}

The second ingredient is that the Schwartz functions are dense in $L^p$ for $p \in \RR$ with $1 \leq p < \infty$.%
\tlink{Analysis/Distribution/SchwartzSpace/Basic.lean\#L1383}
\begin{lstlisting}[mathescape]
theorem SchwartzMap.denseRange_toLpCLM {μ : Measure E} {p : ℝ≥0∞}
    (hp : p ≠ ∞) (hp' : 1 ≤ p) : DenseRange (SchwartzMap.toLpCLM ℝ F p μ)
\end{lstlisting}
This is proved by using that every smooth compactly supported function is a Schwartz function and then proving that every $L^p$ function
can be approximated by smooth compactly supported functions.

Together with an abstract extension argument (sometimes referred to as the BLT theorem~\cite{ReedSimon1}*{Theorem I.7}),%
\tlink{Analysis/Normed/Operator/Extend.lean\#L190}
we arrive at the definition of the Fourier transform for $L^2$ functions%
\tlink{Analysis/Fourier/LpSpace.lean\#L49}
\begin{lstlisting}[mathescape]
def Lp.fourierTransformLI :
    (Lp F 2 (volume : Measure E)) ≃ₗᵢ[ℂ] (Lp F 2 (volume : Measure E))
\end{lstlisting}
where \lstinline{X ≃ₗᵢ Y} denotes an invertible linear isometry and \lstinline{volume : Measure E} is the canonical measure associated to
a finite-dimensional inner product space.

It is an easy consequence that taking the Fourier transform and embedding of various spaces commute:
\begin{lstlisting}[mathescape]
theorem toLp_fourier_eq (f : $\schwartz$(E, F)) : $\FT$ (f.toLp 2) = ($\FT$ f).toLp 2

theorem toTemperedDistribution_fourier_eq
    (f : Lp F 2 (volume : Measure E)) : $\FT$ (f : $\schwartz$'(E, F)) = ($\FT$ f : Lp F 2)
\end{lstlisting}

\subsection{Sobolev spaces}

Finally, we discuss how the tools developed can be used to define Sobolev spaces.
While the Gagliardo-Nirenberg-Sobolev inequality has been formalized in Lean~\cite{vanDoorn2024},
we present the first formalization of Sobolev spaces.

For $k \in \NN$, a tempered distribution $u \in \schwartz'(\RR^d)$ is in the Sobolev space of order $k$ if
\begin{align}\label{eq:Sobk}
    \partial^\alpha u \in L^2 \text{ for all } \abs{\alpha} \leq k
\end{align}
and we then write $u \in H^k(\RR^d)$.

Motivated by the fact that the Fourier transform of a derivative is a multiplication operator, we set
\begin{align}\label{eq:Lambda}
    \Lambda_s \coloneqq \FT^{-1} \ang{\xi}^s \FT\,,
\end{align}
where $\ang{\xi} = (1 + \abs{\xi}^2)^{1/2}$ and $s \in \RR$. In particular, we have that $\Lambda_2 = (1 + (2\pi)^{-2}\Delta)$,%
\footnote{The factors of $2\pi$ appear because the Fourier transform is defined with the phase function $2\pi x \xi$.}
where $\Delta$ denotes the positive Laplacian on $\RR^d$.
The Fourier-theoretic definition of the Sobolev space on $\RR^d$ is then
\begin{align}\label{eq:Sobs}
    H^s(\RR^d) \coloneqq \set{ u \in \schwartz'(\RR^d) \colon \Lambda_s u \in L^2} \,.
\end{align}
It can be shown that for $s = k \in \NN$, this definition agrees with \eqref{eq:Sobk} (cf.~\cite{Hormander1}*{p.~240}).

While a full theory of theory of Sobolev spaces is not yet formalized, we have the following partial results:
we define Fourier multiplier,
that is operators of the form $\FT^{-1} m(\xi) \FT$ for a function $m$ of temperate growth
and the Sobolev space \eqref{eq:Sobs} as unbundled distributions.%
\tlink{Analysis/Distribution/Sobolev.lean\#L149}
As a direct consequence of the Fourier inversion theorem,
we obtain that for $s = 0$ this space is isomorphic to $L^2$.%
\tlink{Analysis/Distribution/Sobolev.lean\#L153}
Since $\abs{\xi}^2 / \ang{\xi}^2$ is bounded,
the Fourier multiplier with symbol $\abs{2 \pi \xi}^2$ maps $H^s$ to $H^{s-2}$,
and we have shown that this is an extension of the Laplacian acting as an operator on $\schwartz' \to \schwartz'$.%
\tlink{Analysis/Distribution/Sobolev.lean\#L344}
In particular, this shows that $\Delta$, initially defined on $\schwartz(\RR^n)$, extends to a map $H^2(\RR^n) \to L^2(\RR^n)$.

\section{Conclusion}
We have described the progress on formalizing abstract functional analysis and distribution theory with the goal of
formally verifying results in partial differential equations.
We define (tempered) distributions and Sobolev spaces, which has been stated as an important goal by various authors~\cites{Boldo2023, vanDoorn2023}.

The example of the Fourier transform on $L^2$ indicates that using Schwartz functions in conjunction with abstract functional analysis
arguments, avoids a lot of calculations with $L^p$ functions. Using $L^p$ functions directly in \mathlib is more tedious since they are not functions, but
equivalence classes of functions. In particular, it is often necessary use the specialized tactic \lstinline{filter_upwards} to prove an
almost everywhere (in)equality from other (in)equalities that only hold almost everywhere.

While it is currently not possible to prove non-trivial results about solutions of partial differential equations,
this work provides an important step towards a theory of formalized partial differential equations.

\end{document}